# DESIGN, MANUFACTURING, ASSEMBLY, AND LESSONS LEARNED OF THE PRE-PRODUCTION 325 MHz COUPLERS FOR THE PIP-II PROJECT AT FERMILAB *

J. Helsper[1,†], S. Wallon[2], D. Passarelli[1], D. Longuevergne[2], S. Kazakov[1], N. Solyak[1]
[1]FNAL, Batavia IL, 60510, USA
[2]Université Paris-Saclay, SNRS/IN2P3, IJCLAB, 91405 Orsay, France

## Abstract

Five 325 MHz high-power couplers will be integrated into the pre-production Single Spoke Resonator Type-II (ppSSR2) cryomodule for the PIP-II project at Fermilab. Couplers were procured by both Fermilab and IJCLAB for this effort. The design of the coupler is described, including design optimizations from the previous generation. This paper then describes the coupler life cycle, including design, manufacturing, and assembly, along with the lessons learned at each stage.

## INTRODUCTION

The pre-production Single Spoke Resonator Type-II (ppSSR2) couplers will provide radio frequency (RF) input to the superconducting accelerating cavities housed within the ppSSR2 cryomodule (CM) [1], which is part of the PIP-II Project [2]. Five ppSSR2 couplers will be used in the ppSSR2 CM string. Nine ppSSR2 couplers were procured; five being procured by FNAL, and four being procured and contributed in-kind by IJCLAB [3]. The ppSSR2 couplers are predated by the prototype Single Spoke Resonator Type-I (pSSR1) couplers [4], which were successfully used and tested on the pSSR1 CM [5]. The ppSSR2 couplers share many design details with the prototype High Beta 650 MHz (pHB650) couplers, which are also part of the PIP-II Project [6].

## DESIGN

The critical design components of the ppSSR2 coupler are shown in Fig. 1.

A single alumina ceramic window separates the beamline volume from atmosphere. The window is brazed to pliable copper sleeves which allow for thermal expansion without undue stress. The copper antenna contains an internal stainless steel (SS) tube which provides cooling air and additional stiffness necessary for transportation and handling. TiN coating was applied to several of the alumina windows, and the results are discussed separately [7]. A high voltage (HV) bias of 3.5 kV is maintained during operation between inner conductor/antenna and the outer conductor, resulting in suppressed multipacting activity. The antenna tip is symmetric and fixed, and so Qext is not adjustable.

* WORK SUPPORTED, IN PART, BY THE U.S. DEPARTMENT OF ENERGY, OFFICE OF SCIENCE, OFFICE OF HIGH ENERGY PHYSICS, UNDER U.S. DOE CONTRACT NO. DE-AC02-07CH11359.
† jhelsper@fnal.gov

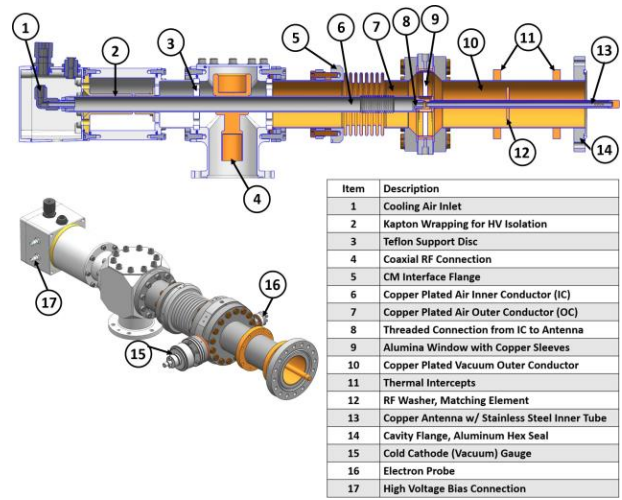

| Item | Description |
|---|---|
| 1 | Cooling Air Inlet |
| 2 | Kapton Wrapping for HV Isolation |
| 3 | Teflon Support Disc |
| 4 | Coaxial RF Connection |
| 5 | CM Interface Flange |
| 6 | Copper Plated Air Inner Conductor (IC) |
| 7 | Copper Plated Air Outer Conductor (OC) |
| 8 | Threaded Connection from IC to Antenna |
| 9 | Alumina Window with Copper Sleeves |
| 10 | Copper Plated Vacuum Outer Conductor |
| 11 | Thermal Intercepts |
| 12 | RF Washer, Matching Element |
| 13 | Copper Antenna w/ Stainless Steel Inner Tube |
| 14 | Cavity Flange, Aluminum Hex Seal |
| 15 | Cold Cathode (Vacuum) Gauge |
| 16 | Electron Probe |
| 17 | High Voltage Bias Connection |

Figure 1: The full ppSSR2 coupler assembly.

The antenna assembly, cold outer conductor (OC), air inner conductor (IC), and air OC were all designed to be furnace brazed. The cold OC is plated with 12 microns of OFHC copper to reduce RF losses.

The DC Block, shown in Fig. 2, serves to isolate the coupler input and protect the RF amplifier from HV bias.

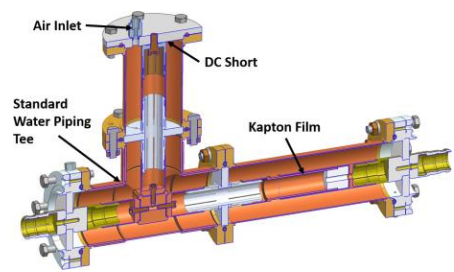

Figure 2: DC Block Design

### Changes from Previous Design

The pSSR1 coupler design [4] is shown in Fig. 3. The following is a summary of the major design changes implemented on the ppSSR2 couplers: air inlet tube changed from removable to fixed, air outer conductor has a single bellow and is copper plated, the ceramic disk stress relief sleeves were lengthened, the antenna flange uses ConFlat ® sealing instead of 'diamond' seals, the E-Probe was moved to the vacuum OC flange instead of the tube, a vacuum gauge was added, the vacuum OC wall thickness was increased to improve robustness, and copper plating was added to the vacuum OC. While most of these changes were inspired by the lessons learned from pSSR1 couplers, while others came from the 650MHz couplers [8] [9].

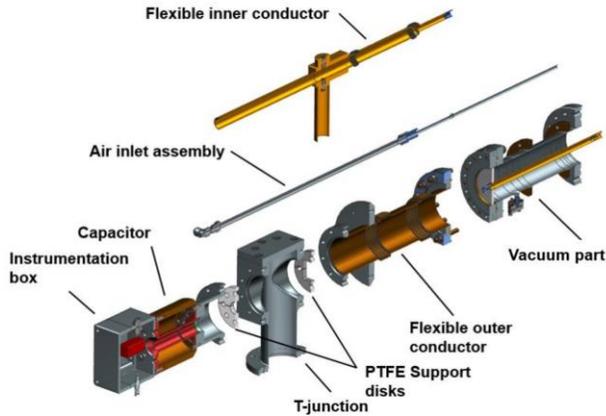

Figure 3: pSSR1 Coupler Design [4]

*Analysis*

The RF, thermal, and structural analyses for the ppSSR2 couplers was satisfactory, and comparable to that made for the pSSR1 couplers [10]. The results of this analysis were separately confirmed by IJCLAB and their vendor using AN-SYS®. The RF and thermal analyses were only preformed to verify the heat load to the cavity, and utilized a 3D-shell model of the vacuum side fixed to the cavity. Evaluation of the radiated power to the cavity, using ANSYS® with diffuse radiation hypothesis, found the heat loads very similar to the original values calculated by FNAL with CST®.

## MANUFACTURING

The couplers procured and delivered by IJCLAB were based on FNAL's drawing set and manufacturing specification. IJCLAB and FNAL would regularly communicate during the manufacturing and inspection process, and IJCLAB provided all oversight and was the primary design authority over their couplers. Only major design discrepancies from IJCLAB's couplers were referred to FNAL for disposition, and otherwise IJCLAB operated largely independent of FNAL.

*FNAL Experience*

To improve the delivery schedule, the release for manufacturing was split into two phases: the first being an approval to procure 'piece parts', only requiring approval of vendor drawings, and the second being an approval to manufacture couplers, requiring all quality control (QC) samples and documentation to be completed. Per FNAL's drawings, it was the vendor's responsibility to select the proper brazing tolerances, and while not required, it was suggested the vendor should redesign any brazed joints deemed sub-optimal (this was not done by FNAL's vendor). The vendor did not rely on any sub-contractors except for the procurement of machined piece parts and bellows.

FNAL staff visited their vendor late in the manufacturing process of the first couplers to inspect the units. Excessive braze material was found in the ConFlat ® seal of the antenna assembly, determined to be a result of erroneous powdered braze filler from a neighboring braze joint. Additionally, the inner braze from the antenna tip to the SS tube was not made, as the vendor thought it would induce buckling due to the dissimilar contraction rate of OFHC Copper and SS. This potential issue was determined to be less severe than the potential effects of the un-brazed tip (low stiffness, microphonics issues) and so the braze was made. Other issues were found, including nicks and scratches to sealing surfaces, but overall, the coupler quality was satisfactory.

*IJCLAB Experience*

**Changes to Processes and Interfaces** IJCLAB's vendor modified brazed joints based on their vast prior experience. Tolerances and clearances were solely determined by the vendor. Since individual parts have unavoidable dimensional variations, IJCLAB and the vendor pursued individual pairings of male and female parts to minimize potential braze material overflow.

Even though the vendor is specialized in metal and metal-ceramic brazing, and owns a large range of brazing furnaces, some alternative solutions to brazing were chosen. Electron beam welding (EBW) was used to connect the brazed antenna to the TiN coated ceramic window sub-assemblies since TiN can't tolerate brazing temperatures. Several air parts were laser welded including the flange-tube connections and the bellows-tube connections. TIG welding was used for small flange-tube connections on the VAC OC. These changes were done to reduce cost and avoid delays since the vendor's furnaces were occupied with other large scale production jobs.

These changes required modification to the relevant interface designs, shown in Fig. 4. For the ceramic outer sleeve, extra material was added as a shoulder, which served to stiffen the part (better tolerances) and was better served to EBW. This design change would likely work well on a brazed assembly as well. One flange on the air side required an undercut next to the weld area so the welded parts would be of similar effective thickness.

**Antenna Window Assembly** One significant challenge during manufacturing was achieving the antenna straightness requirement of <0.5 mm radial with respect to the ideal axis of the flange. Specialized equipment was made which worked in conjunction with a coordinate measurement machine (CMM) to straighten antennas. It was only used on one of the spare antennas, and with limited success. The device will be tested further by IJCLAB. The required antenna straightness was later revised to be less stringent once the potential impact to Qext was confirmed to be minimal.

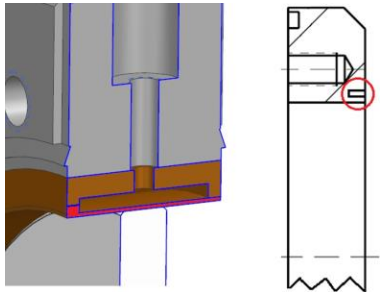

Figure 4: Left: Update to window sleeve. Right: Update to Vessel Flange

**Schedule Delays due to Reliance on Sub-Contractors**
Machining was outsourced to several sub-contractors, along with two for copper plating (one for tubes/assemblies, and one specialized in bellows), and one for cleaning and packaging. This "chain" of sub-contractors was the weak point in the schedule, and caused significant delays. Additional schedule delays came from the change of Project Manager mid-project, and COVID-19 had a negative impact on economic revival and delivery of sub-contractors.

Copper plating of the bellows proved difficult due to variations in plating thickness, process tuning issues, repeatability, and the inability to verify thickness on the produced parts.

### Lessons Learned

The lessons learned during the manufacturing stage are as follows: allowing the vendor to purchase 'piece parts' after drawing review but before sample generation proved successful; solid braze filler should be used in place of powdered; flanges must have protective plastic covers in place as much as possible; parts to be brazed can be paired individually after dimensional checks to prevent braze overflow; reliance on sub-contractors can lead to schedule and QC variability out of the vendors direct control; some sub-contractors will be willing to improve delivery times, and as a side effect, quality may suffer; vendors should make a visit to sub-contractors to witness first steps of work and set out basic rules and procedures (e.g. using clamp mounting instead of a three jaw chuck); in person visits are crucial to finding issues in couplers; antenna straightness required in drawings should accurately reflect needs for RF performance and not be unnecessarily strict.

## INCOMING INSPECTION

### FNAL Couplers

The quality of the air side coupler components, shown in Fig. 5, was excellent, with very few issues found. The copper plating was adherent and free of defect, the brazes were of good quality, and individual parts conformed to the relevant drawings and specifications. The only major issue found was that the bellows outer conductors had inconsistent length due to varying levels of bellow compression/elongation. This issue was remedied via incremental stretching of the bellow and re-inspection of copper plating. Stretching was necessary as the bellow outer conductor interfaces with the vacuum vessel, and this interface location can't be significantly moved. The inspection took place in an open-air workspace and followed written procedures.

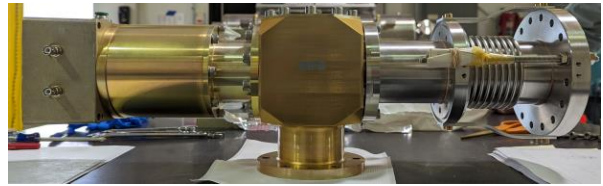

Figure 5: Air Side Coupler Assembly

The quality of the vacuum side coupler components proved more problematic. The first two vacuum side units did not meet the specified cleanliness requirements, which was caused by a misunderstanding on the part of the vendor. Of the five couplers, two vacuum side outer conductors were found to have under-filled braze joints, which only had a blackened appearance at first. Prodding with thin copper foil (soft enough not to scratch, but firm enough to remove loose material) revealed that the defect was nearly half the depth of the brazed joint and 3-4 mm in width. Given the defect location (beamline vacuum space, facing into the cavity) the units were sent back for repair. It is thought that inspection prior to copper plating was insufficient. Additionally, discussion with the vendor found the length of the braze joint (12 mm) and flow orientation (cavity facing side was the end of the flow path) to be non-optimal. The inspection of vacuum side components took place in an ISO 4 clean room, and followed written procedures.

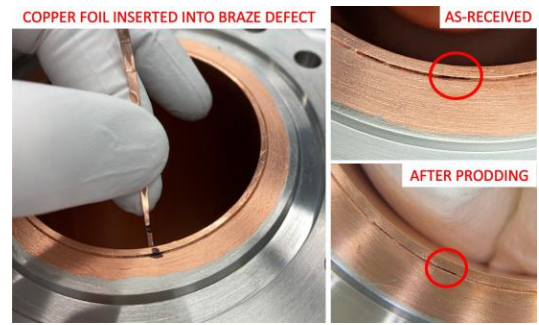

Figure 6: Outer Conductor Brazing Defect, Cavity Flange

On one pair of antennas, foreign black material was found embedded in the surface of the antenna tips, shown in Fig. 7. A Scanning Electron Microscope found the material to be consistent with stainless steel. The vendor successfully removed the material, and it is hypothesized that the material adhered to the antenna during electropolishing (EP) due to remnant particulates in the tank. Such an occurrence had not been seen by the vendor in all their experience manufacturing couplers. Additionally, the external surface finish of the antennas was sub-optimal, determined to be caused by FNAL's design and use of standard ½" copper piping for the primary antenna body.

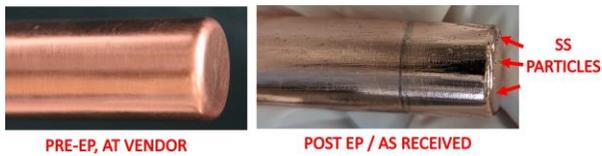

Figure 7: Contamination at Antenna Tip

Only two of the antennas were found to be visibly bent, thought to be caused by the dissimilar contraction rate noted by the vendor. As the radial position of the antenna tip has minimal effect to the Qext of the SSR Cavities, no remedy was necessary. While not cause for return, some small dents were seen on the inside of the outer conductor wall, below the copper plating. Discussion with the vendor revealed that as FNAL specified a given tube to be used, a finishing pass couldn't't be made while also maintaining the specified inner diameter. Other issues included minor damage to sealing surfaces and braze material adjacent to the ConFlat ® sealing surface which could prevent UHV sealing.

All other aspects of the vacuum side components met or exceeded expectations, including, but not limited to, shipping configuration, copper plating quality, general brazing quality, cleanliness (upon return to FNAL), and dimensional control. An accepted antenna and Vacuum OC can be seen in Fig. 8.

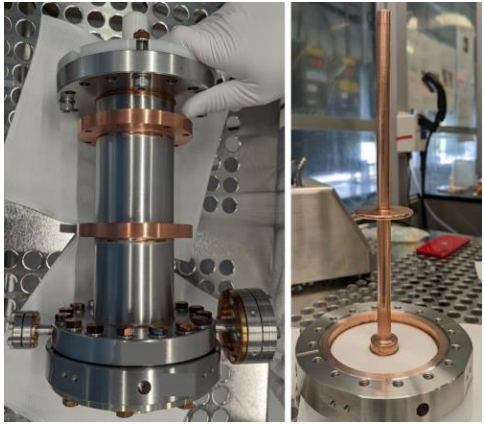

Figure 8: FNAL Accepted Components

### IJCLAB Couplers

All couplers procured by IJCLAB were shipped directly to FNAL for incoming inspection. Any issues that required remediation were fixed by FNAL due to the logistical complexities and scheduling issues involved in their return to IJCLAB's vendor. At the time this was written, only two vacuum side assemblies had been received and inspected, and so there is no commentary on the air side assemblies.

One of the coupler's antenna was found to be significantly bent, as shown in Fig. 9. As the stiff protective cover was also bent, it is hypothesized that the fully assembled and bagged coupler was dropped at the vendor's sub-contractor and not disassembled for inspection afterwards. The coupler passed leak check, and will be conditioned as-is on the test stand to determine any possible effects the damage has on RF performance.

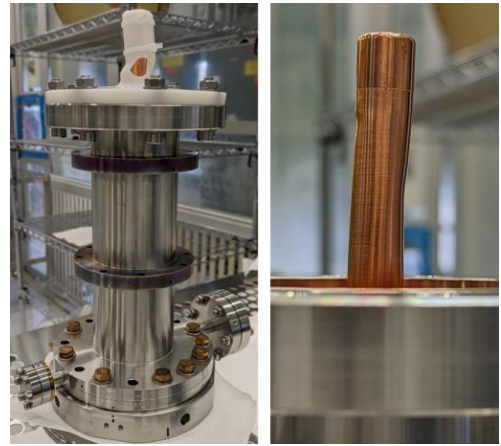

Figure 9: Damaged Antenna

The overall quality of EBW and brazing was excellent. The exterior antenna surface exceeded expectations as the vendor had machined the antenna tube from solid stock instead of using the tube specified in the original design. The copper plating quality and lack of oxidation met expectations. Due to several small instances of damage to sealing surfaces, many of the components had to be re-cleaned. However particle counting found the UHV areas of the parts in compliance with requirements. Fig. 10 shows the non-damaged antenna and a Vacuum OC.

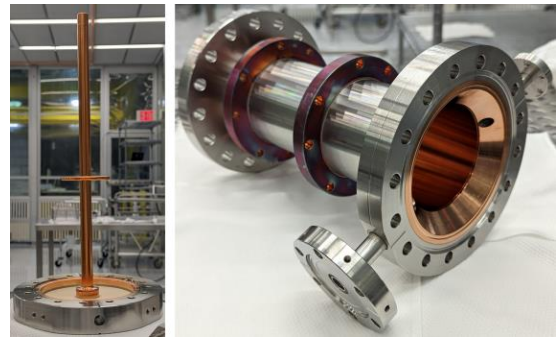

Figure 10: IJCLAB Accepted Components

Several other small issues were found, including the usage of SS bolts with SS nuts (galling prone), ethanol wiping found several bolts to be contaminated, the instrumentation bolt connecting to the window was severely over-tightened, several bolts from the antenna to outer conductor were hand-tight, one threaded hole on the antenna had to be re-tapped, and most Si-Brz bolts were found to be flaking.

### Lessons Learned

The lessons learned from the incoming inspection stage are as follows: bellows assemblies must have better dimensional control; the first coupler units should be inspected immediately prior to shipment and not weeks / months beforehand, final assembly torque specifications must be provided; QC of threaded holes should include fit-checks; the vendor must better optimize brazed joints, 'off the shelf' tubes should not be used for critical components; custom fasteners (Si-Brz) require additional QC; and the vendor should perform a final detailed inspection prior to shipment to ensure conformity to requirements.

## ASSEMBLY, TRANSPORT, AND BAKING

As stated previously all couplers were received, inspected, and tested at FNAL. There are no lessons learned for the remaining sections unique to one vendor or the other.

After components passed visual inspection and particle counting, the assembled couplers were installed to the RF test chamber shown in Fig. 11. When assembled, the antenna has a 3 mm radial gap to the capacitive coupling tube which the antenna is centered within. This assembly is performed by hand with the assistance of two technicians. The clean stand has minimal features near the coupler connection to minimize flow turbulence during assembly. Unfortunately, the very bent antennas could not be used with the current stand components, as they made physical contact with the capacitive coupling tube, preventing the use of HV bias during testing.

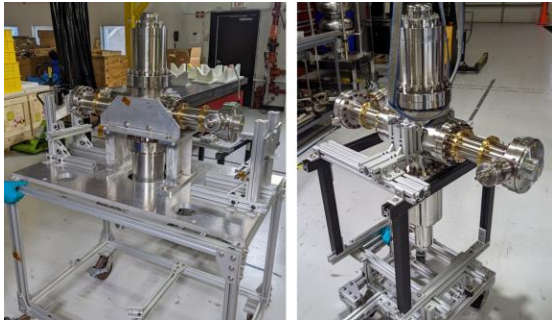

Figure 11: Left: RF Stand, Right: Clean Stand

Once out of the clean room, the chamber assembly is lifted and placed into the RF Stand cart. This is then transported to the onsite baking facilities, where the entire cart assembly is baked at 120 C for 48 hours while under vacuum. Residual gas analysis (RGA) scans are taken before and after 120 C baking to confirm lack of contamination and appropriate partial pressures.

Once transported to the RF testing facility, the air side couplers and RF Coaxial lines are assembled, as shown in Fig. 12. The air side is supported by frames which mimic the cryomodule interface. Discussion of RF testing and the associated results are not within the scope of this article, but can be found in [12].

### Lessons Learned

The lessons learned from clean assembly, baking, and RF assembly are as follows: optimization of the chamber mounting is necessary to aide ease of installation/removal;

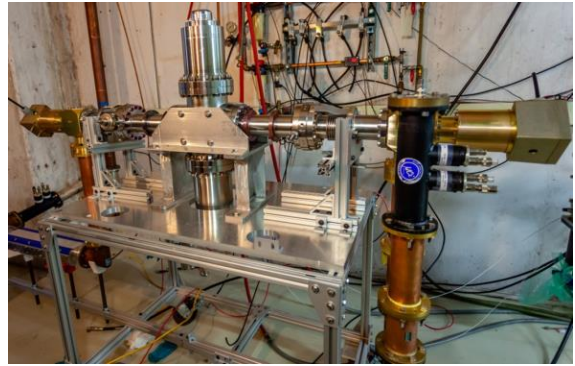

Figure 12: Assembled RF Test Stand

lifting features on the cart would assist with transportation; and larger diameter capacitive coupling tubes are required for the chamber to accommodate bent antennas.

## SUMMARY

In summary, the ppSSR2 couplers procured by both FNAL and IJCLAB met the technical requirements necessary to be used as part of the ppSSR2 CM String. Procurement, manufacturing, QC, assembly, and integration have all provided valuable experiences which will positively influence the production SSR coupler design and specifications, along with the other PIP-II couplers at FNAL.

While the issues listed may seem disparaging of either vendor, the authors would like to point out that the manufacture and delivery of couplers is no easy task, especially for small quantity orders when a steady supply chain cannot be established. Challenges are frequent and numerous producing these small batches of couplers, and the performance of the couplers [12] serves as a record of each vendor's ability to deliver high quality RF couplers.

## ACKNOWLEDGEMENTS

The work completed would be impossible without Fermilab's dedicated technicians R. Kirschbaum, T. Fermanich, D. Plant, and their supervisor D. Bice.